\documentclass[showpacs,preprintnumbers,amsmath,amssymb]{revtex4}

\usepackage{graphicx}
\usepackage{subfigure}
\usepackage{dcolumn}
\usepackage{bm}

\begin{document}

\title{Relaxation to equilibrium of the expectation values in macroscopic quantum systems}
\author{Takaaki Monnai}%
\email{monnai@suou.waseda.jp}%
\affiliation{$*$Department of Applied Physics, Osaka City University,
3-3-138 Sugimoto, Sumiyoshi-ku, Osaka 558-8585, Japan}
\begin{abstract}
A quantum mechanical explanation of the relaxation to equilibrium is shown for macroscopic systems for nonintegrable cases and numerically verified.
The macroscopic system is initially in an equilibrium state, subsequently externally perturbed during a finite time, and then isolated for a sufficiently long time.    
We show a quantitative explanation that the initial microcanonical state typically reaches to a state whose expectation values are 
well-approximated by the average over another microcanonical ensemble.    
\end{abstract}
\pacs{05.70.Ln,05.40.-a}
\maketitle
\section{Introduction}
It is our experience that isolated thermodynamic systems would eventually relax to a stationary state called equilibrium\cite{Zwanzig,Deutsch}.
In thermodynamics, this statement is regarded as requirements for the relevant choice of the model and its macroscopic descriptions.
Recently, it is shown that equilibrium states are very popular in the extremely large Hilbert space of state vectors for macroscopic quantum systems\cite{Lebowitz,Lebowitz2,Lebowitz4,Tumulka,Popesucu,Popesucu2,Reimann,Reimann2,Tasaki,Sugita}. 
The expectation values with respect to a randomly sampled pure state are typically well described by equilibrium ensembles\cite{Lebowitz,Reimann,Sugita}.
And it is expected that initial nonequilibrium state approaches to typical state of equilibrium. 
Then it is reasonable to explore a dynamical explanation for the macroscopically observed behavior of the relaxation to equilibrium\cite{Lebowitz2,Lebowitz4,Popesucu2}.
Indeed the coarsgrained macroscopic quantities typically yield a decomposition of the total Hilbert space with a subspace of overwhelming dimension which represents equilibrium\cite{Lebowitz2,Lebowitz4}.
Based on this property, arbitrary state approaches and stays near equilibrium\cite{Lebowitz2,Lebowitz4,Tumulka}, which is a modern version of quantum ergodic theorem\cite{vonNeumann}.
By assuming the thermodynamically normal fluctuation of the quantities, the approach to equilibrium can be derived\cite{Tasaki}.   
Also the reduced density matrix of huge isolated system approaches to time averaged state and stays near equilibrium for almost all the time\cite{Popesucu2}.
Description of the relaxation is also important in the context of nonequilibrium theorems\cite{Jarzynski1,Jarzynski2}, which relies on the use of thermodynamic ensembles.
As for the isolated system, the relaxation is not guaranteed for the total density matrix itself. Indeed the total density matrix does not actually relax to a thermal state in the course of unitary time evolution.
Linear phonons in the initial microcanonical state which are externally perturbed and excites only few modes will not approach to a microcanonical state.       
For classical systems, there exist integrable systems which do not reach an equilibrium state even after a long waiting time due to the conserved quantities.
For isolated quantum systems, there are always conservative quantities such as projection operators on a specific energy eigenstates\cite{Popesucu2,Sugita,Reimann2,Rigol}. 
And it has been investigated which conservative quantities would affect the thermalization property\cite{Rigol4}.
The thermalization of a few body quantities has been established for some nonintegrable systems whose energy spectrum do not tend to degenerate\cite{Rigol,Rigol2}.
There are also reports that the reduced equilibrium ensemble average works for spin chains both for integrable and nonintegrable parameters\cite{Saito}, which is consistent with the typicality of thermal state\cite{Lebowitz} but in marked contrast to \cite{Rigol,Rigol2} at least for the small system sizes.
It is possible that there are formulations of relaxation properties which are not very sensitive to the nonintegrability.
Since we assume the initial microcanonical state, thermalization for the expectation values of the conservative quantities is trivial for autonomous systems as explained later on.
Also the evaluation of off-diagonal elements for the nonintegrable case shows quantitatively good agreement with the numerical values, which is a sufficient condition to guarantee the relaxation property.

We insist that from the uniformly random sampling assumption which holds for nonintegrable cases and the initial equilibrium state, the expectation values of quantities which polynomially depend on the system size typically show relaxation to the equilibrium values.
Especially, it is pointed out that the off-diagonal elements of the density matrix does not contribute to the expectation values.
In this article, we explain how the statement is made quantitative for the time-dependent unitary time evolution.
And the explanation is numerically verified for a spin chain with a magnetic field in the nonintegrable regime. 
This paper is organized as follows.
In Sec.II, we present our model with the precise external forcing procedure.
In Sec.III, the relaxation of the expectation values are explained.
Sec.IV is devoted to a summary.      
\section{Macroscopic systems}
Let us consider a macroscopically large but finite initially isolated system. 
After initial time $t=0$, an external forcing acts on the system, and the total Hamiltonian $H(t)$ depends on time.  
Until $t=0$, the density matrix describing the initial state is assumed to be microcanonical specified by an energy $E$
\begin{equation}
\rho(0)=\frac{1}{\Omega(0)}\delta(E-H(0)),\label{canonical}
\end{equation}
where $\Omega(0)$ is the density of the states. 
In the course of time evolution, the external work is done on the system through the time dependence of $H(t)$.

The deterministic external forcing acts during the time interval $0\leq t \leq T_0$, and switched off for $t>T_0$. 
It is expected that after a sufficiently long waiting time, i.e. at $t={\it T}\gg T_0$, the density matrix reaches a state which yields approximately the same expectation values as the microcanonical ensemble $\rho({\it T})=\frac{1}{\Omega({\it T})}\delta(E+\Delta E-H({\it T}))$ with an energy shift $\Delta E$.
\section{Relaxation of the expectation values}
It is our statement that under the uniformly random sampling assumption valid for the nonintegrable cases, the microcanonical ensemble well approximates actual expectation value of quantity $A$ whose maximum eigenvalue polynomially depends on the system size     
\begin{equation}
{\rm Tr}U\rho(0)U^+ A\cong{\rm Tr}\rho({\it T})A. \label{unitary}
\end{equation}
This evaluation shows that the relaxation to equilibrium is explained as the property of the expectation values instead of the density matrix.
Note that the initial microcanonical state guarantees that Eq.(\ref{unitary}) holds for the conservative quantities with $\rho(T)=\rho(0)$ for autonomous systems irrespective of the integrability.
An important class of observables is extensive quantities which are sums of local quantities such as total magnetization in a spin chain. 
Such quantities mutually almost commute, since the local quantities at spatially different points commute.  

The actual final state is reached by a unitary time evolution
\begin{eqnarray}
&&U\rho(0)U^+ \nonumber \\
&=&e^{-\frac{i}{\hbar}H({\it T})(T-T_0)}U(T_0)\rho(0)U(T_0)^+e^{\frac{i}{\hbar}H({\it T})(T-T_0)} \nonumber \\
&=&\sum_{n,m} c_{n,m} |E_n\rangle\langle E_m|, \label{evolution}
\end{eqnarray}
where we presented the eigenstate of $H(T)$ as $H(T)|E_n\rangle=E_n|E_n\rangle$, $c_{n,m}=e^{-\frac{i}{\hbar}(E_n-E_m) (T-T_0)}\langle E_n|U(T_0)\rho(0)U(T_0)^+|E_m\rangle$, and $U(T_0)=T\{e^{-\int_0^{T_0}\frac{i}{\hbar}H(t)dt}\}$.
After $t=T_0$, the total system is isolated, and evolves by the Hamiltonian $H(T)$.  
On the other hand, for the nonintegrable systems the off-diagonal elements of the physical quantity $A$ is negligible compared to the diagonal elements:
\begin{itemize}
\item[i)] The off-diagonal matrix element $\langle E_n|A|E_m\rangle$ of the nonintegrable system is typically negligible in the macroscopic limit.
Physically, this would be reasonable since it means that transition amplitudes between macroscopically different states due to the perturbation $A$ are extremely small\cite{vanKampen}.
We shall give a quantitative explanation of this statement.
The analysis is based on the high-dimensionality of the Hilbert space as in Refs.\cite{Lebowitz,Popesucu,Reimann,Sugita}. 
Let us diagonalize the quantity $A$ as 
\begin{equation}
A=\sum_n A_n |A_n\rangle\langle A_n|,
\end{equation}
and define its square root 
\begin{equation}
B=\sum_n \sqrt{A_n} |A_n\rangle\langle A_n|.
\end{equation}
Here the spectrum $\{A_n\}$ is assumed to be nonnegative. Without loss of generality, the operators bounded below such as energy and number of particles can be made into this form.  
We also define state vectors $\{|\Phi_n\rangle=B|E_n\rangle\}$ so that the matrix element is expressed as the inner product
\begin{equation}
\langle E_n|A|E_m\rangle=\langle\Phi_n|\Phi_m\rangle. \label{offdiagonal}
\end{equation}
The states $\{|\Phi_n\rangle\}$ are chosen from the extremely large Hilbert space ${\cal H}$ with various directions.  
Thus we assume that the sequence of the normalized vectors $\{\frac{|\Phi_1\rangle}{\sqrt{\langle\Phi_1|\Phi_1\rangle}},\frac{|\Phi_2\rangle}{\sqrt{\langle\Phi_2|\Phi_2\rangle}} ,...\}$ is regarded as a {\it uniformly random sampling} from ${\rm dim}{\cal H}-1$ dimensional unit sphere as we will confirm for the case of nonintegrable spin chain.    
It is then straightforward to show that the mean square of the inner product is smaller than $\frac{\|A\|^2}{{\rm dim}{\cal H}}$\cite{Lebowitz}, which we will show later.
Here $\|A\|$ is the maximum of the eigenvalues of $A$. 

Let us derive the inequality for the inner product
\begin{equation}
\langle|\langle\Phi_n|\Phi_m\rangle|^2\rangle\leq\frac{\|A\|^2}{{\rm dim}{\cal H}}, \label{inner}
\end{equation}  
where the bracket shows the average with respect to the uniform random sampling of $\frac{|\Phi_n\rangle}{\sqrt{\langle\Phi_n|\Phi_n\rangle}}$ from the unit sphere.
Uniform random sampling is expressed by  the vector representation of $\frac{|\Phi_n\rangle}{\sqrt{\langle\Phi_n|\Phi_n\rangle}}$ in an orthogonal complete basis as 
$\vec{d}_n=(\cos\theta_1,\sin\theta_1\cos\theta_2,\sin\theta_1\sin\theta_2\cos\theta_3,...,\sin\theta_1\sin\theta_2\cdot\cdot\cdot\sin\theta_{d-1})$ with $d={\rm dim}{\cal H}$.
Hereafter we use the abbreviated notation for the dimension $d$.
The angles $\theta_i$ are uniform random variables on the real axes.
We are interested in particular cases where $\|A\|$ polynomially depends on the system size.
On the other hand, the dimension ${\rm dim}{\cal H}$ grows exponentially as the system size increases.
Then inequality (\ref{inner}) shows that the off-diagonal elements Eq.(\ref{offdiagonal}) is extremely small for the macroscopic system.
We numerically show that Eq.(\ref{inner}) actually holds and hence random sampling assumption is reasonable for a quantum spin chain.
The spin chain is regarded as a nice example, since it provides observables, which is a sum of the local quantities.  
The Hamiltonian of the spin chain is chosen as
\begin{equation}
H=-J\sum_{j=1}^{N}\sigma_j^z\sigma_{j+1}^z+\alpha\sum_{j=1}^N\sigma_j^x+\gamma\sum_{j=1}^N\sigma_j^z,
\end{equation}
where $J$ stands for the exchange interaction between neighboring sites, and a constant magnetic field ${\bf B}=(\alpha,0,\gamma)$ is applied.
$\sigma_j^i$ is the $i=x,y,z$ component of the Pauli matrix at the site $j$. $\gamma$ controles the nonintegrability.
As a macroscopic quantity, we consider the square of the $x$ component of the total magnetization
\begin{equation}
A=(\sum_{j=1}^N\sigma_j^x)^2. \label{quantity}
\end{equation}
Other choices of $A$ are possible, but Eq.(\ref{quantity}) guarantees a well-defined square root $B=\sum_{j=1}^N\sigma_j^x$.
For example, if we choose $A$ as $x$ component of the total magnetization or $A=(\sum_{j=1}^m\sigma_j^x)^2$ $m\leq N$, we still observe behavior resembles to Fig.1.
\begin{figure}
\center{
\includegraphics[scale=0.8]{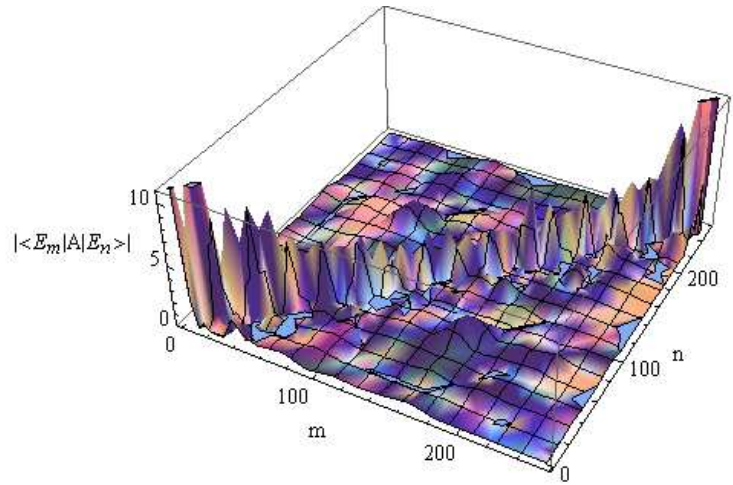}
}
\caption{The absolute values of matrix elements $|\langle E_m|A|E_n\rangle|$ of $A=(\sum_{j=1}^N\sigma_j^x)^2$. The average of the off-diagonal elements is the same order as the average of the diagonal elements divided by $\sqrt{d}$. The system size is $N=8$. We have confirmed qualitatively similar results for $6\leq N\leq 11$.}
\end{figure}
\begin{figure}
\center{
\includegraphics[scale=0.8]{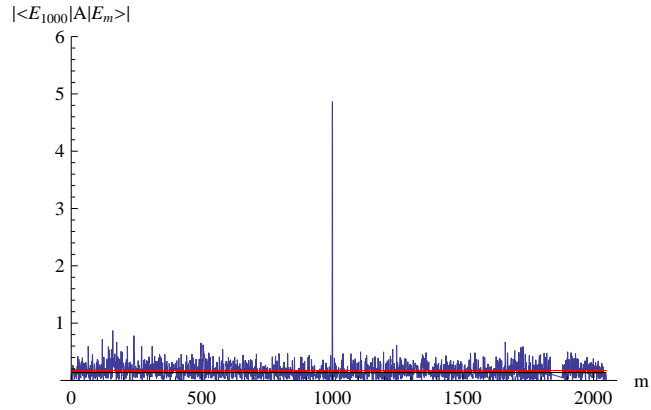}
}
\caption{The absolute values of matrix elements $|\langle E_{1000}|A|E_n\rangle$ of $A=(\sum_{j=1}^N\sigma_j^x)^2$. $y$ and $z$ components give similar
 results. The red line shows average of the off-diagonal elements $\langle|\langle E_{1000}|A|E_n\rangle|\rangle=0.173$, and the black line gives the average of the diagonal elements divided by $\sqrt{2^N}$, $0.138$. These two lines are mutually very close.}
\end{figure} 
The system sizes $6\leq N\leq 11$ are explored. Note that the dimension of the total Hilbert space is $2^N$ and can be large for relatively small $N$.
By diagonalizing the Hamiltonian matrix $H$ under the periodic boundary condition, numerical eigenvectors $|E_n\rangle$ are obtained, and the absolute values of all the matrix elements $|\langle E_m|A|E_n\rangle|$ are shown  in Fig.1.
The data are shown only for parameters $J=1$, $\alpha=1$, and $\gamma=0.5$, however, we have confirmed similar behavior for various values of $\gamma\geq 0.1$. 
To verify that Eqs.(\ref{inner},\ref{product}) are satisfied quantitatively, we also show the matrix elements $|\langle E_{1000}|A|E_n\rangle|$ with a fixed value of $m=1000$ for $N=11$.
The statistical mean value of the off-diagonal elements agrees with the theoretical estimation $\frac{\langle A\rangle}{\sqrt{d}}\cong\frac{\|A\|}{\sqrt{d}}$ as  shown in Fig.2.
\begin{figure}
\center{
\includegraphics[scale=0.8]{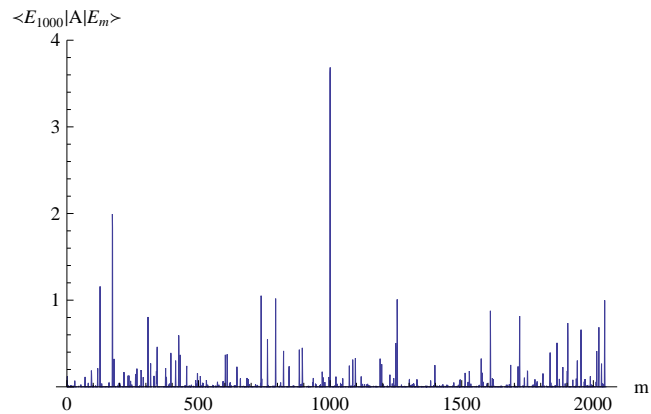}
}
\caption{The absolute values of matrix elements $|\langle E_{1000}|A|E_n\rangle|$ of $A=(\sum_{j=1}^N\sigma_j^x)^2$. $y$ and $z$ components give similar results.
The data are shown for parameters $J=1$, $\alpha=1$, and $\gamma=0$, where the Hamiltonian is integrable. The mean of off-diagonal elements is $0.03719$, and the theoretical value is $\frac{|\langle E_{1000}|A|E_{1000}\rangle|}{\sqrt{2^N}}=0.1068$.}
\end{figure} 
This estimation is considered as a generic property of macroscopically different states for the nonintegrable systems, since it is expressed only by the typical value of $A$ and the dimension $d$.  
Thus for large $d=2^N$, the directions of $|\Phi_n\rangle$ are orthogonal to each other, and are considered to be distributed on the $d-1$ dimensional unit sphere without any bias to a specific direction.
\begin{equation}
\frac{\langle|\langle\Phi_m|\Phi_n\rangle|^2\rangle}{\sqrt{\langle\Phi_m|\Phi_m\rangle\langle\Phi_n|\Phi_n\rangle}}=\frac{1}{d} \label{product}
\end{equation}
 is immediately derived, since the expectation values of each component are the same and their sum should be unity.
Since $|E_n\rangle$ and $|E_m\rangle$ are normalized, 
\begin{eqnarray}
&&\langle \Phi_n|\Phi_n\rangle\langle\Phi_m|\Phi_m\rangle \nonumber \\
&=&\langle E_n|A|E_n\rangle\langle E_m|A|E_m\rangle \nonumber \\
&\leq&\|A\|^2. \label{normalization}
\end{eqnarray}
The inequality (\ref{inner}) derives from Eqs.(\ref{product},\ref{normalization}).    
For comparison, the matrix elements $|\langle E_{1000}|A|E_n\rangle|$ at $N=11$ for the integrable case $\gamma=0$ are shown in Fig.3.
As in Fig.2, $A=(\sum_{j=1}^m \sigma_j^i)^2$ is explored.
For $m\leq N$ and $i=y,z$, $|\langle E_{1000}|A|E_n\rangle$ behaves as in the nonintegrable cases.
But for $i=x$, the smallness of the off-diagonal elements is affected.  
Aside from the dominant contribution from the diagonal element, there are subdominant contributions as shown in Fig.3.
The same tendency takes place also for $m\leq N$.
This shows that the uniformly random sampling assumption is affected at least for the relatively small system size.

From Eq.(\ref{inner}), the coefficient $c_{n,m}$ of Eq.(\ref{evolution}) is safely replaced by $\langle E_n|U\rho(0)U^+|E_m\rangle \delta_{n,m}$ in the evaluation of ${\rm Tr}U\rho(0)U^+A=\sum_{m,n}\langle E_m|U\rho(0)U^+|E_n\rangle\langle E_n|A|E_m\rangle$.
The off-diagonal elements of $U\rho(0)U^+$ does not contribute to the expectation value of the quantity $A$, since $|\langle E_m|A|E_n\rangle|$ $(m\neq n)$ is neglected..
More quantitatively, the error caused by neglect of the off-diagonal elements is indeed negligible by random phase approximation, since there are $d^2$ terms of $O(\frac{1}{d^2})$ with various phases.  
\item[ii)] The diagonal elements of the density matrix $U\rho(0)U^+$ is well-approximated by those of a microcanonical state with respect to $H(T)$.
It is important to note that the initial microcanonical state specifies an energy scale $E$.  
Indeed the diagonal elements are expressed as 
\begin{eqnarray}
&&\langle E_n|U\rho(0)U^+|E_n\rangle \nonumber \\
&=&\frac{1}{\Omega(E(0))}\langle E_n|\delta(UH(0)U^+-E)|E_n\rangle \nonumber \\
&=&\frac{1}{\Omega(E(0))}\sum_m \delta(\tilde{E}_m-E)|\langle E_n|\tilde{E}_m\rangle|^2, \label{element}
\end{eqnarray}
where we introduced the normalized eigenstates of $UH(0)U^+$ as $UH(0)U^+|\tilde{E}_n\rangle=\tilde{E}_n|\tilde{E}_n\rangle$, i.e. $U|E_n(0)\rangle=|\tilde{E}_n\rangle$ with an eigenstate of $H(0)$, $|E_n(0)\rangle$.
The set of states $\{|E_n\rangle\}$ and $\{|\tilde{E}_n\rangle\}$ are related by a unitary transformation $\sum_n|E_n\rangle\langle\tilde{E}_n|$.
In the presence of perturbation, $|\langle E_n|\tilde{E}_m\rangle|^2=|\langle E_n|U|E_m(0)\rangle|^2$ would be non negligible only when the conservation of the energy is well-satisfied after the long waiting time, 
i.e. $E_n\cong \tilde{E}_m+\Delta E$ with the energy change $\Delta E$ from the initial to final times caused by external perturbation during $0\leq t\leq T_0$.

We show a quantitative estimation of $\Delta E$. The matrix element $\langle E_n|\tilde{E}_m\rangle$ is evaluated as $\delta_{nm}+\frac{\langle\tilde{E}_m|H(T)|\tilde{E}_n\rangle}{\tilde{E}_m-\tilde{E}_n}(1-\delta_{nm})$ up to the first order of the perturbation $H(T)-UH(0)U^+$, 
where the factor $\langle\tilde{E}_m|H(T)|\tilde{E}_n\rangle$ would be of order $\sqrt{\frac{\|H(T)\|^2}{d}}$ as in Eq.(\ref{inner}).
Then it is immediately shown that $\tilde{E}_m$ contributes to $\langle E_n|\tilde{E}_m\rangle$ only when $\tilde{E}_m$ and $\tilde{E}_n$ are sufficiently near $|\tilde{E}_m-\tilde{E}_n|\leq\frac{E}{\sqrt{d}}$, otherwise the ratio $\frac{\langle\tilde{E}_m|H(T)|\tilde{E}_n\rangle}{\tilde{E}_m-\tilde{E}_n}$ is negligible.
Such $|\tilde{E}_m\rangle$ is expanded as $|\tilde{E}_m\rangle=\sum_n d_{n,m}|E_n\rangle$, where the coefficient $d_{n,m}$ is non negligible for $\tilde{E}_n$ sufficiently near $\tilde{E}_m$.
The energy change $\Delta E$ is determined as 
\begin{eqnarray}
&&\Delta E=\langle \tilde{E}_m|(H(T)-UH(0)U^+)|\tilde{E}_m\rangle \nonumber \\
&=&\sum_k |d_{km}|^2E_k-\tilde{E}_m \nonumber \\
&\cong&E_n-E,
\end{eqnarray}
which is almost independent of the suffix $n$.
Here we evaluated as $\sum_k|d_{k,m}|^2E_k\cong E_m\cong E_n$ and $\tilde{E}_m=E$ from the Dirac delta in Eq.(\ref{element}). 
The important property used here is a continuity of the mapping, i.e. when $|\tilde{E}_n-\tilde{E}_m|$ is small enough compared to $\sqrt{\frac{E^2}{d}}$, $|E_n-E_m|$ is also sufficiently small.
   
Therefore the third line of Eq.(\ref{element}) has a sharp peak at $E_n=E+\Delta E$ as a function of $E_n$ and is proportional to the function $\delta(E_n-E-\Delta E)$ as  
\begin{eqnarray}
&&\langle E_n|U\rho(0)U^+|E_n\rangle \nonumber \\
&\cong&\frac{1}{\Omega(T)}\delta(E_n-E-\Delta E) \nonumber \\
&=&\langle E_n|\frac{1}{\Omega(T)}\delta(H(T)-E-\Delta E)|E_n\rangle, \nonumber \\
\end{eqnarray} 
where the density of the states at $t=T$ is determined uniquely from the normalization, and $H(T)|E_n\rangle=E_n|E_n\rangle$ is used. 
The diagonal elements is thus given by the microcanonical ensemble. 
\end{itemize}
Then as far as the expectation value is concerned, the state $U\rho(0)U^+$ should be well-described by the microcanonical ensemble $\rho(T)$.

\section{Summary}
In conclusion, by assuming that the initial state is prepared as a microcanonical ensemble and the uniformly random sampling assumption holds as in the nonintegrable systems, the actual state at time $t=T$ can be replaced by another microcanonical ensemble in the evaluation of quantities which polynomially depend on the system size.  
The derivation is based on the high dimensionality of the Hilbert space and consequent uniformly random sampling assumption valid for the nonintegrable systems, and restrictions for the system size dependence of observables as well as the perturbative treatment of unitary transformation of the energy eigenstates.
The validity of the uniformly random sampling assumption for $\frac{|\Phi_n\rangle}{\sqrt{\langle\Phi_n|\Phi_n\rangle}}$ is numerically verified for a nonintegrable spin chain.
It is also remarked that the initial state can be out of equilibrium, i.e. the initial density matrix is $\rho(0)=U_0\frac{1}{\Omega(0)}\delta(E-H(0))U_0^+$ with a unitary transformation $U_0$, since this can be regarded as the state evolved from an actual initial state $U_0^+\rho(0)U_0=\frac{1}{\Omega(0)}\delta(E-H(0))$.
\section{Acknowledgment}
The author is grateful to Professor A.Sugita for fruitful discussions.
This work is financially supported by JSPS program Grant in aid 22$\cdot$7744.

\end{document}